\documentclass[%
 reprint,
unsortedaddress,
 amsmath,amssymb,
 aps,
]{revtex4-2}

\usepackage{graphicx}
\usepackage{dcolumn}
\usepackage{bm}
\usepackage{amssymb}
\usepackage[utf8]{inputenc}
\usepackage[T1]{fontenc}
\usepackage{mathptmx}
\usepackage{etoolbox}
\usepackage{bbm}
\usepackage{xcolor}
\usepackage{hyperref}

\begin{document}

\title[Local and global optimization in Parallel Minority Games]{Local and global optimization in Parallel Minority Games}

\author{Soumyajyoti Biswas}%

\affiliation{ 
Department of Physics, SRM University - AP, Andhra Pradesh 522240, India
}%

\author{Jnanesh Yaramati, Kavya Bellamkonda, Krishna Rastogi, Devesh Chaudhary}

\affiliation{ 
Department of Computer Science and Engineering, SRM University - AP, Andhra Pradesh 522240, India
}

\date{\today}

\begin{abstract}
The Parallel Minority Game (PMG) refers to a set of Minority Games (MG), played in parallel, where each agent only has two choices to pick from, but each choice can host agents of many kind i.e., their other alternative can be from any other choices. While the pay-off function remains the same as that in the MG -- agents picking the less crowded of their two choices win positive pay-off -- the optimization of resource allocation is significantly harder in the PMG. While a global optimization demands a uniform population in all choices, a local optimization attempts to balance the population in the two choices for a given agent. In the MG these two objectives coincides, but generally in the PMG these are competing. We study several non-dictated, stochastic strategies and compare their efficiencies in attaining the local and global optimization objectives. Counterintuitively, a strategy with partial information of populations perform the best in terms of population fluctuation and overall payoff maximization. 
\end{abstract}

\maketitle

\section{Introduction}
The Minority Game (MG) is a well studied model in optimization of resource allocation \cite{mg_rev}. Proposed as a modification of the El Farol Bar problem \cite{elfarol}, the MG deals with two choices that $N=2M+1$ agents must choose between. The objective for each player is to remain in the minority of the two options in their repeated and independent choices, for most of the time. Of course, a complete random choice gives a fluctuation between the populations of the two choices that scale as $\sqrt{N}$. However, it was shown that when agents have a past memory of their choices and outcomes, it is possible to reach a level of population fluctuation using deterministic strategies that are significantly less than the random choices \cite{mg1}. 

Later on, it was shown that if the agents adopt a stochastic strategy, then the minimum possible fluctuation (Populations at $M$ and $M+1$) could be reached in a very short time ($\sim \log \log (N)$) \cite{dhar}, however an additional information of the population difference must be available to the agents. It was shown subsequently that the low level of population fluctuation could be achieved even without that information, if the agents simply make a guess of the population of the other location and assume it to decrease with time (following an annealing schedule) \cite{pre}. The time required to reach the low fluctuation state then scales as $\log N$. These strategies follow the stochastic strategies proposed in the Kolkata Paise Restaurant problems \cite{kpr1,kpr2}.

The Parallel Minority Game (PMG) \cite{amit} is a generalization of the MG, where there are many ($M$) choices, even though each agent can only choose between two of those. This means, in a given location, there could be various types of agents, in the sense that their alternate location could come from any one of the remaining $M-1$ choices. The minimum fluctuation state is difficult to achieve here, as this is essentially a coupled set of MGs, played in parallel. The increase or decrease in population in one location may affect the minority/majority status at many other locations. A similar effort along the lines of the stochastic strategy was mentioned above was done for PMG. It was shown that in order to avoid a sudden overcrowding at a given location could only be avoided when the agents are supplied with a time delayed information of the population of their alternate location \cite{ankith}. However, the question remains that the level of information available to the agents could substantially affect the final configuration and effective strategies of the agents. 

In this work, we systematically study the effect of the level of information available to the agents in finding the optimal population configuration and pay-off maximization of the agents. The various levels of information available can significantly affect the strategies, and thereby change the optimal state. It turn out, rather counterintuitively that having the highest level of information does not necessarily give the most optimized strategy for higher pay-off or the minimization of the population fluctuation.

\section{Methods}
The objective of resource utilization in the case of PMG has two components, that can be competing against each other. On one hand, the individual agents want to remain in the minority of their two choices (the traditional MG objective) and on the other hand the population fluctuation among all the choices should be minimized. Note that in MG, the two objectives coincide. In that case, the condition where most number of people can stay in the minority is the same condition as reducing population difference (hence fluctuation) between the two choices. But in the case of PMG, the conditions need not align for all agents. 

The non-dictated strategies of the agents in optimizing their pay-off probability can take various forms. In particular, as discussed in ref. \cite{ankith}, that a version of stochastic strategy can lead to low fluctuations in populations of various choices. However, there are two main questions that need addressing: first, the best performing strategy was the one where the agents had the information of population of their current choice at time $t$ and they also had the information of population at their alternate choice at time $t^{\prime}<t$, where $t^{\prime}$ is the time when the agent last visited that alternative location. Using that strategy, it was shown that the populations reach a frozen state, similar to a spin-glass like behavior, where no further change in choices happen. However, it is clear that the state is not the most optimal. Therefore, it is important to investigate if it is possible to use an annealing technique, similar to simulated annealing in spin glasses, where the agents can reach a better optimal state. The second issue is regarding the information available to the agents at all times. In the traditional MG, the only information the agents have access to is whether they are in the minority or the majority at any given time (and the memory of that). In the stochastic strategy proposed in \cite{pre}, the information of the population was also required as an additional information. However, it was also showed that one can use an annealing schedule, i.e. an exponential or linear decrease of a guessed value of the population, and use the same stochastic strategy to reach the optimal state in a short ($\log N$) time. Therefore, it is important to explore the effect of an annealing schedule for the PMG with different levels of information available with the agents. 

We study systematically the effect of having different levels of information with the agents and the effect of the annealing method. Below we detail the methods:

\begin{figure*}[tbh]
\centering
  \includegraphics[width=0.45\linewidth]{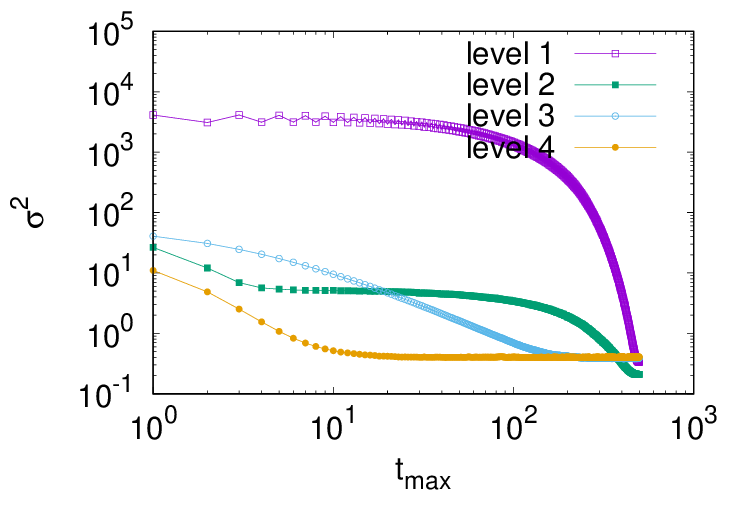}
  \includegraphics[width=0.45\linewidth]{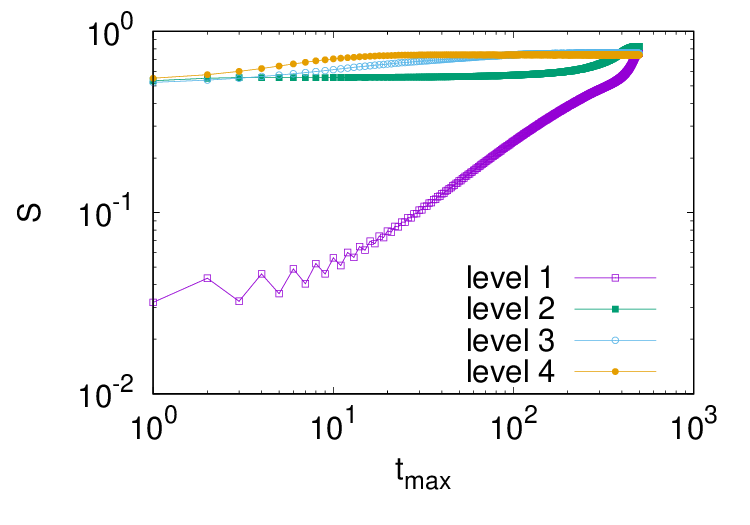}
\caption{The left hand side figure shows the time variation of population fluctuation for various strategies (levels of information), and the right hand side figure shows the fraction of agents in their minority locations for the same strategies. It is seen that $\sigma^2$ is minimized and $S$ is maximized for level 2, which is not the highest level of information available. Here, $t_{max}=500$, $g=101$ and $M=100$.}
\label{all_S_sigma}
\end{figure*}

Note that the strategies are arranged according to the level of information shared with the agents. In each level, the information of the earlier levels are also assumed to be present.
\begin{itemize}
    \item (level 1) The agents only know if they are in the majority or minority between their two choices. In this case, the agents switch to their alternate location with a probability 
    \begin{equation}
        p_i(t)=(1-t/\tau_i),
    \end{equation}
    where $\tau_i$ are taken from a uniform distribution in a range $(\tau_{min},\tau_{max})$ unless otherwise stated. 

    \item (level 2) The agents know the population of their current location at the current time ($n_i^x(t)$). They flip only if they are in the majority (between their own two choices). The population at the alternate location (say, $n_i^y(t)$) is not known to the agent. But it is known that $n_i^y(t)<n_i^x(t)$ (when they are in the majority). So, they make a reasonable guess that  $\tilde{n}_i^y=n_i^x(t)-\sqrt{n_i^x(t)}(1-t/\tau_i)$. The flip probability is then
    \begin{equation}
        p_i(t)=\frac{n_i^x(t)-\tilde{n}_i^y(t)}{2n_i^x(t)}.
    \end{equation}

     \item (level 3) The agents know the global average of the population in all locations $g=N/M$. Here also the agents flip with a probability (when they are in the majority and the population at current location is higher than the global average) given by
     \begin{equation}
         p_i(t)=\frac{n_i^x(t)-g}{2n_i^x(t)}.
     \end{equation}

     \item Finally (level 4), when the agents have the information for the populations of the current and the alternate locations, they flip with a probability (only when in the majority) given by
     \begin{equation}
         p_i(t)=\frac{n_i^x(t)-n_i^y(t)}{2n_i^x(t)}.
     \end{equation}
     
\end{itemize}

We simulate the model using Monte Carlo simulations, and find the pay-off and population fluctuations for each of the above cases. 

\begin{figure*}[tbh]
\centering
  \includegraphics[width=0.45\linewidth]{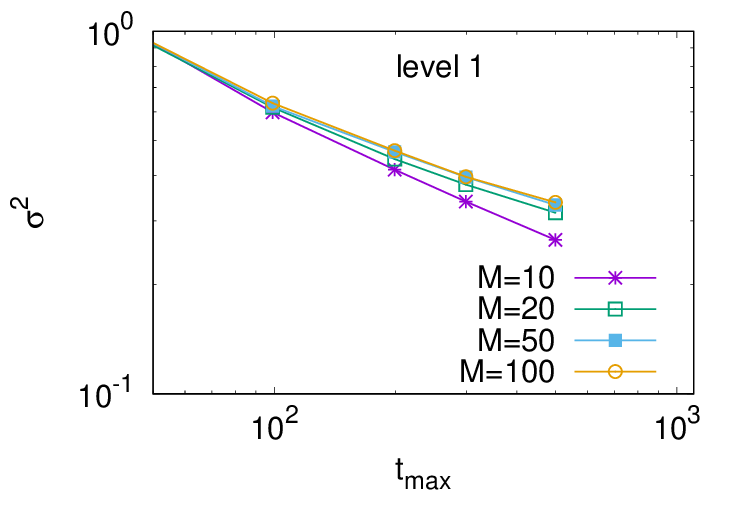}
  \includegraphics[width=0.45\linewidth]{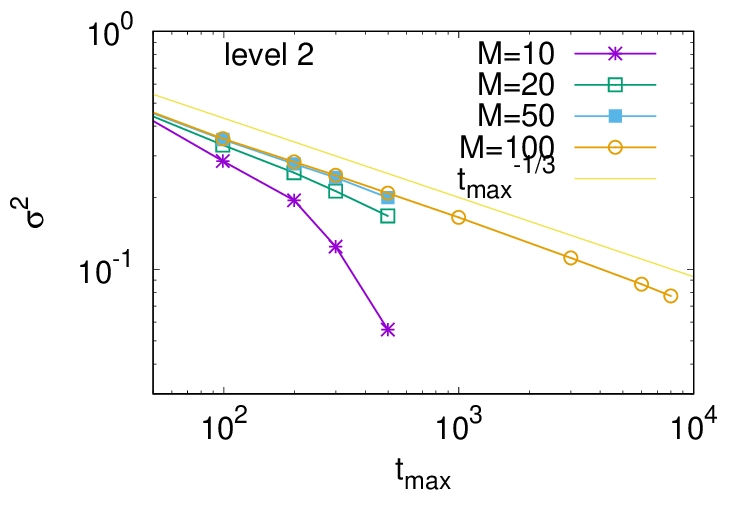}
  \includegraphics[width=0.45\linewidth]{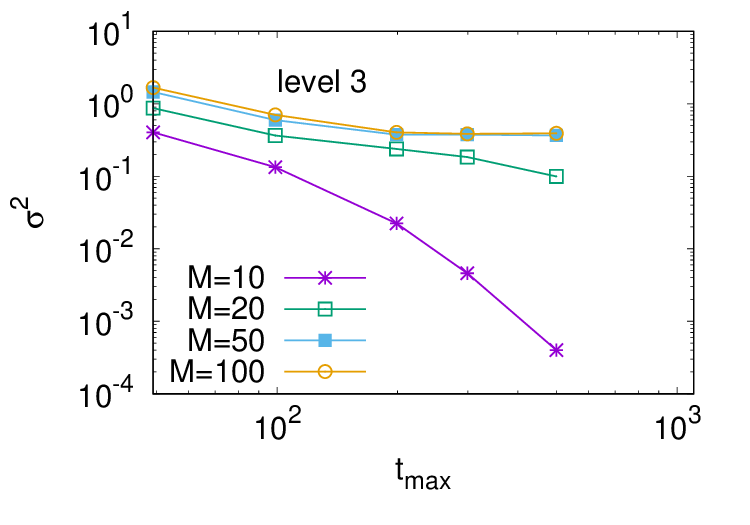}
  \includegraphics[width=0.45\linewidth]{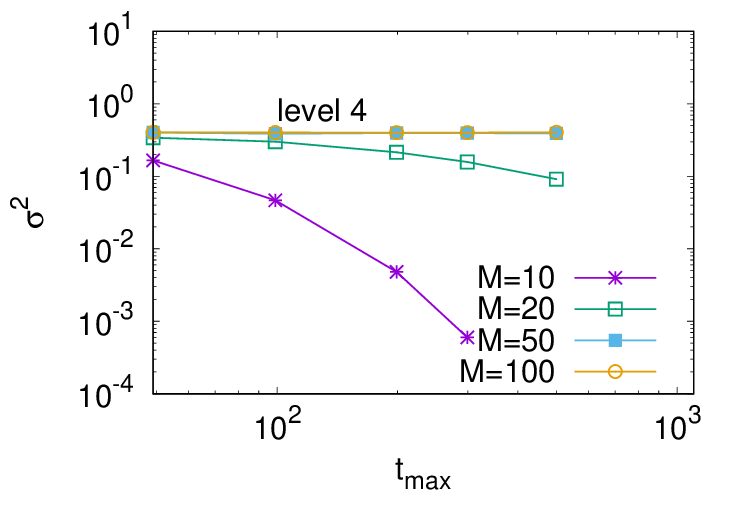}
\caption{The variation of the population fluctuations $\sigma^2$ are shown for various levels of information as a function of $t_{max}$. For all levels, the fluctuation either saturates for large values of $M$ or show a tendency to saturate, except for level 2. In that case it continuous to decay in a power law, following $t_{max}^{-1/3}$. }
\label{all_M}
\end{figure*}

\section{Results}
In the simulations, we keep $g=M/N=101$ fixed, but vary $M$ (hence $N$), and measure two quantities: the population fluctuation $\sigma^2=\frac{1}{M}\sum\limits_i(n_i-g)^2$, where $n_i$is the population at the $i-th$ location. We also measure the fraction of agents $S$ who are in the minority location between their two choices. 

In Fig. \ref{all_S_sigma}, we show the variation of the population fluctuation $\sigma^2$ as a function of time ($t$), where each time step refers to one attempted update of each agent's location. As mentioned before, the agents only attempt to switch their locations with a probability $p$, when they are in the minority of their two choices. Note that in a given locations, some agents may be in the minority while the others can be in the majority, as their alternate locations (and therefore its population) might be different. We see that in level 2 the population fluctuation is minimized, which is not the highest level of information available. Also note that the simulation time runs until $t=t_{max}$, which is the maximum annealing time-scale chosen for an agent. Of course, for agents whose $\tau_i$ values are less than $t_{max}$, the switching probability $p=0$ for $t>\tau_i$.

In Fig. \ref{all_M}, we show the variations of the end point value (value at $t=t_{max}$) of the population fluctuation for various $M$ and for different strategies. It is seen that as $M$ is increased, the $\sigma^2$ values saturates (or tend to saturate) for all strategies except for level 2. In that case, until the limit of $t_{max}$ values studied, the fluctuation decreases, following a power-law behavior $\sigma^2\sim t_{max}^{-1/3}$. This makes the counter-intuitive result that lowest population fluctuation (and highest pay-off) possible. It also shows that at least for the end-point values of population fluctuation and pay-off, the two optimization conditions align well for level 2 strategies.

\section{Discussions and conclusions}
The optimal population distribution that minimizes the fluctuation is difficult to find in the PMG, as it is essentially a coupled set of MGs running in parallel \cite{aryan}. Earlier efforts have shown that a delayed time information produced a better optimization than when the current information (regarding populations) are provided to the agents. 

Here we see that even a much lower level of information, where the agents only have information about their current location, performs the best in terms of the population fluctuation as well as the overall payoff to the agents. However, the time required to reach the optimal state is divergent, as the fluctuation decays in a power-law with the maximum value of annealing schedule time. 

In the case of the MG, the level of information was similar, when the optimal distribution was obtained in $\log N$ time i.e., the agents only knew of their current location's population. But given that in PMG, there are multiple locations and agents in a given location can have their alternate locations distributed (uniformly) among those, it does not automatically provide the information about their alternate location's population. However, if that information is included, the population fluctuation becomes higher, due to possible overcrowding at the low population locations. 

In conclusion, the population fluctuation (global optimization) and payoff maximization (local optimization) in the PMG do not follow a monotonic variation with the level of information available to the agents. A strategy with partial information is better equipped in finding both the global and local optimal state than that with full information of the populations available to the agents.

\end{document}